# Towards Hardware implementation of video applications in new telecommunications devices

Lamjed Touil, Abdessalem Ben Abdelali, Mtibaa Abdellatif, and Elbey Bourennane.

**Abstract**—among the areas, most demanding in terms of calculation is the telecommunication and video applications are now included in several telecommunication devices such as set-top boxes, mobile phones. Embedded videos applications in new generations of telecommunication devices need a processing capacity that can not be achieved by the conventional processor, to work around this problem the use of programmable technology has a lot of interest. First, Field Programmable Gate Arrays (FPGAs) present many performance benefits for real-time image processing applications. The FPGA structure is able to exploit spatial and temporal parallelism. In this paper, we present a new method for implementation of the Color Structure Descriptor (CSD) using the FPGA circuit. In fact the (CSD) provides satisfactory image indexing and retrieval results among all color-based descriptors in MPEG-7. But the real time implementation of this descriptor is still having problems. In this paper we propose a method for adapting this descriptor for possible implementation under the constraints of the video processing in real time. We have verified the real-time implementation of the (CSD) with an image size of 120*80 pixels.

——————————— ◆ ———————————

## 1 INTRODUCTION

Image processing is widely used in many applications, including industrial, medical imaging manufacturing, and security systems. And real-time image processing is difficult to realize on a serial processor. This is due to many factors such as the large data set represented by the image, and the complex operations which may need to be performed on the image. At real-time video rates of 25 frames per second a single operation performed on every pixel of a 768 by 576 colour image (PAL frame) equates to 33 million operations per second. The large quantity and flow of information submitted by the video stream, and the constraints imposed by real-time applications require the solution purely hardware. The use of reconfigurable architectures based FPGA circuits is of great interest since they have the advantage of greater flexibility, with integrated advanced resources and the possibility of parallelization important treatment due to their structure.

An FPGA offers a compromise between the flexibility of general purpose processors and the hardware-based speed of ASICs. Performance gains are obtained by bypassing the fetch-decode-execute overhead of general purpose processors and by exploiting the inherent parallelism of digital hardware. In recent years this circuit has resulted in an increasing interest in their employ as implementation platforms for image processing applications, principally real-time video processing [1]. And many image segmentation algorithms have already been proposed. However the majority of these algorithms is implemented in software, and has large complexity so that hardware implementation is difficult [2].

Among video applications, we cite MPEG-7 visual descriptors. MPEG-7 is an ISO standard for describing meta-data on multi-media objects, including for instance video, sound and pictures. MPEG-7 visual descriptors record statistics of images and video sequences in color, texture shape of objects and motion. It can be employed in video indexing and retrieval, content summarization, surveillance, etc. These techniques are actually exploited in constrained applications such as set top box, systems and personal video recorders (PVR).

Color is one of important visual attributes for human vision and image processing. In fact color is the most distinguishing visual features in image and video retrieval. It is robust to changes in the background colors and is independent of image size and orientation [3]. Many forms of color distributions and representations are adopted in MPEG-7[4]. In MPEG-7, CSD and SCD provide descriptors are used provide better image indexing and retrieval results [5, 6]. Operational analysis of software simulation for CSD is shown in table 1, explains why CSD

- *Lamjed Touil, Laboratoire EμE, Faculté des Sciences FSM. Monastir, Tunisia.*
- *Abdessalem ben Abdelali, Laboratoire EμE, Faculté des Sciences FSM. Monastir.*
- *Abdellatif Mibaa, Laboratoire EμE, Faculté des Sciences FSM. Monastir.*
- *Elbey Bourennane, Laboratory LE2I University of Burgundy, Dijon, France.*

can not be applied to real time products without the hardware accelerator. And there is no good solution at present [7].

The proposed architecture uses multi access-embedded memories. Recently, technology development has enabled the integration of embedded memories with large storage capacities and large memory-access bandwidth in FPGA. In this paper, we propose a hardware architecture and analysis of the CSD.

The rest of this paper is organized as follows: in section 2, we first describe briefly the algorithm of CSD. Section 3 focuses implementation details of each functional block and then each block design. An example of hardware architecture for this descriptor is presented in this section. In Section 4 we explain different experimentations, and the evolution results of various application modes of the CSD.

TABLE 1
MIPS MEMORY BANDWIDTH OF CSD GENERATOR [2]

| Operation | 1fps | | 3fps | |
|---|---|---|---|---|
| | Number of instructions (MIPS) | Memory bandwidth (Mbytes) | Number of instructions (MIPS) | Memory bandwidth (Mbytes) |
| HMMD | 5.625 | 3.585 | 168.750 | 107.550 |
| Accumulation | 143.657 | 202.456 | 4309.710 | 6073.680 |
| Quantization | 0.051 | 0.001 | 1.517 | 0.039 |
| Others | 0.990 | 0.697 | 29.713 | 20.9801 |
| **Total** | **150.323** | **206.739** | **4509.690** | **6202.170** |

## 2 SPECIFICATION OF THE MPEG-7 COLOR STRUCTURE DESCRIPTOR (CSD)

The color structure descriptor represents an image or an image region by its color distribution. The main function of this descriptor is image to image matching and its intended use is for still-image retrieval. It provides satisfactory image indexing and retrieval results among all color-based descriptors in MPEG-7 standard [8, 9]. The superiority comes from the consideration of space distribution of colors. The CSD expresses the local color structure in an image using an 8x8 structuring element. In fact, instead of characterizing the relative frequency of individual image samples with a particular color, this descriptor characterizes the relative frequency of structuring elements that contain an image sample with a particular color.

The CSD is identical in form to a color histogram but it is different in term of semantics. Suppose the number of colors represented in a histogram is denoted by M; that is, the colors in the image are quantized into M different colors c0, c1, c2, …, cM-1. The color structure histogram can be denoted by h(m), m=0, 1, …,M-1, where the value in each bin represents the number of structuring elements in the image containing one or more pixels with color "Cm ". The final CSD is represented by 1D array of 8-bit quantized values. The performance of this descriptor increases considerably while using the HMMD color space [10].

The Color Structure descriptor shall be defined using the HMMD color space. The color pixels of incoming images in any other color space shall be converted to the HMMD color space and requantized appropriately before extracting the structure histogram. The CSD is defined using four color space quantization operating points: 256, 128, 64, and 32 bins [11]. The number of quantization levels constitutes also the number of the histogram bins. The CSD containing 256 bins is directly extracted from the image based on a 256-cell quantization of the HMMD color space. For a bin number less than 256, the bins can be computed based on a unification of the bins of the 256-bin descriptor.

The Color Structure histogram accumulation step is illustrated by the example of Figure 1. The Histogram is computed by visiting all locations in the image, observing which colors are presented in it, and then updating color bins by adding one, no matter how many the same color pixels exist. The increase (or not) of the bins is determined by the presence (or absence) of the corresponding colors, and not by the count of the pixels of each present color. Therefore, in any given position of the structuring element, the increase can be either 0 or 1. For example, suppose we have 8 different colors and the structuring element has a size of 8 by 8 pixels, the structuring element contains some pixels with color c1, some pixels with color c3 and some pixels with color c6. Then, the bin labeled c1, the bin labeled c3 and the bin labeled c6 would each be incremented once. So, in this location, the Color Structure histogram is incremented three times, once for each color present in the structuring element area.

The spatial extent of the structuring element depends on the image size but the number of samples in the structuring element is held constant by subsampling the image and the structuring element at the same time. The number of samples in the structuring element is always 64, and the distance between two samples in this pattern increases with the image size.

A structuring element with 64 samples is used. While the number of samples is fixed, the spatial extent of the structuring element is adjusted to the image size.

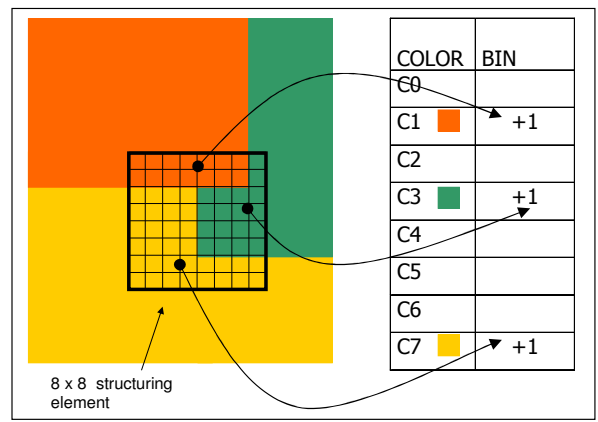

Figure 1    Color histogram accumulation [2].

The additional information about color structure makes the descriptor sensitive to certain image features to which the color histogram is blind [9]. The CSD is extracted from an image in the HMMD color space [9]. The accumulated bin values are normalized by the number of locations of the structuring element and non-linearly quantized. The histogram size is variable and may be 32, 64, 128 or 256. As the standard allows different histogram sizes, it also proposes an approach for resizing descriptors of different size to make them comparable.

The CSD is identical in form to a color histogram but is semantically different. Specifically, the CSD is a one-dimensional array of 8-bit quantized values,

$$CSD = \bar{h}_s(m), \quad m \in \{1, \ldots, M\}$$

Where M is chosen from the set {256, 128, 64, 32} and where s is the scale of the associated square structuring element. The matching procedure determines the similarity of two visual items by computing the similarity between their color-structure histograms [9]. The matching procedure determines the similarity of two visual items by computing the similarity between their color-structure histograms. Let $h_A(i)$ be the histogram vector of visual item A, and $h_B(i)$ be the histogram vector of visual item B, the norm histogram distance is given by:

$$dist(h_A, h_B) = \sum_{i=1}^{N} |h_A(i) - h_B(i)|$$

The CSD is identical in form to a color histogram but it is different in term of semantics. Suppose the number of colors represented in a histogram is denoted by M; that is, the colors in the image are quantized into M different colors c0, c1, c2, …, cM-1. The color structure histogram can be denoted by h(m), m=0, 1, …,M-1, where the value in each bin represents the number of structuring elements in the image containing one or more pixels with color "Cm ". The final CSD is represented by 1D array of 8-bit quantized values. The performance of this descriptor increases considerably while using the HMMD color space [9]. Figure 2 shows the bloc diagram of entire system.

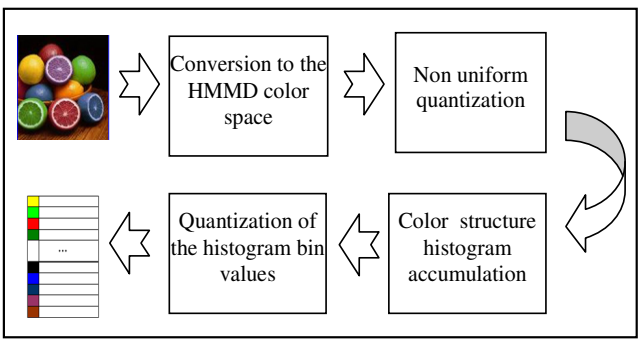

Figure 2   The extraction steps of the color structure descriptor

## 3   FUNCTIONAL ARCHITECTURE OF THE PROPOSED SYSTEM

To validate the specification of the CSD detailed in the previous paragraph, we propose the following block diagram illustrated in the figure 3.

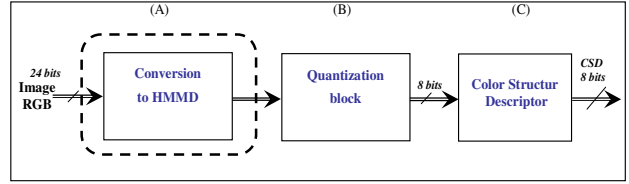

Figure 3   Bloc diagram

Block diagram consists of 3 independent parts (A, B, C).

➢ The block (A) convert the RGB signal to a HMMD signal.

➢ The block (B) is a quantization block.

➢ The block (C) gerates the CSD histogram

The results presented in Table 1, show that the classical solution implementation of the CSD algorithm is not feasible; subsequently we propose our solution for hardware validation of this descriptor. We begin with the RGB to HMMD converter

### 3.1. Presentation of the RGB/hmmd converter

The representation of color is a very important step in the processing and image compression, in particular, it plays a fundamental role in the loss of image compression and image retrieval. MEG-7 supports three color spaces, HSV, the matrix linear and hmmd. The HSV color space and hmmd are made of non-linear transformations, they have better results in applications for search and retrieval and they are closely related to human perception of color.

The color space hmmd composed of black, white, color and hue. There are five attributes in the color space hmmd: Hue, Max, Min, Diff and Sum.

- Max: max(R, G, B); indicates how much black color the image has, giving the flavour of shade or blackness;

- Min: min(R, G, B); indicates how much white color the image has, giving the flavour of tint or whiteness;

- Diff: Max-Min; indicates how much gray the image contains and how close to the pure color, giving the flavour of tone or colorfulness;

- Sum: (Max+Min)/2; simulates the brightness of the color.

The transformation from RGB to hmmd is a nonlinear reversible transformation. The conversion is done according to the following approach:

This module has a parallel structure to improve the processing speed. It needs of total 29 ohrloge cycles to complete converting operation. The comparators find the maximum and the minimum among RGB. The subtractors calculate the difference with red and green, green and blue, blue and red, and the maximum and minimum. The multipliers multiply these values by a predefined constant. After that these multiplied values are divided by the dividers. Figure 6 shows the bloc diagram of RGB to HMMD converter module.

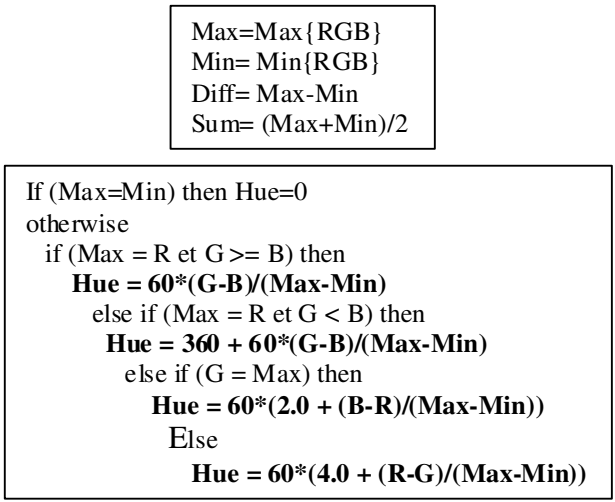

This module has a parallel structure to improve the processing speed. It needs of total 29 ohrloge cycles to complete converting operation. The comparators find the maximum and the minimum among RGB. The subtractors calculate the difference with red and green, green and blue, blue and red, and the maximum and minimum. The multipliers multiply these values by a predefined constant. After that these multiplied values are divided by the dividers. Figure 6 shows the bloc diagram of RGB to HMMD converter module.

This architecture has three layers of parallelism:

➢ The first layer that determines the output sum

➢ The second one that determines the output Diff

➢ The third, which determines the output Hue

## 3.2. Quantization bloc.

Figure 4 shows a slice of the HMMD space in the diff-sum plane for zero hue angle and depicts the quantization cells for the 256-cell operating point, it is shows in tableau 2. Cut-points defining the subspaces are indicated in the figure by vertical lines in the color plane. The diff-axis values that determine the cut-points are shown in black at the top of the dashed cut-point markers along the upper edge of the plane. Horizontal lines within each subspace depict the quantization along the sum-axis. The quantization of hue angle is indicated by the gray rotation arrows around each cut-point marker. The gray number to the right of a rotation arrow corresponds to the number of levels to which hue has been quantized in the subspace to the right of the cut-point, for example, the bijective mapping between color-space cells and descriptor bin indices is given explicitly by the numbers within the cells. Figure 5 shows the block quantization. The entries in this block (Hue, Diff, Sum, Min and Max) are derived from the RGB / HMMD block

This block uses a set of comparators to deliver the output information (HMMD), this block uses quantization levels to determine the range of each value

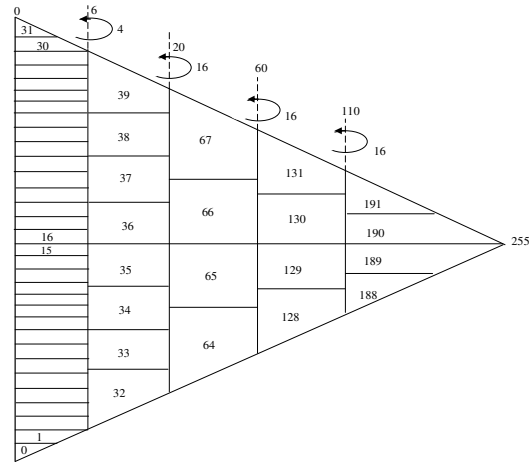

Figure 4    Slice of the HMMD space

TABLE 2.

HMMD COLOR SPACE QUANTIZATION FOR COLOR STRUCTURE DESCRIPTOR

| Subspace | Number of quantization levels for different numbers of histogram bins | | | | | | | |
|---|---|---|---|---|---|---|---|---|
| | 256 | | 128 | | 64 | | 32 | |
| | Hue | Sum | Hue | Sum | Hue | Sum | Hue | Sum |
| 0 | 1 | 32 | 1 | 16 | 1 | 8 | 1 | 8 |
| 1 | 4 | 8 | 4 | 4 | 8 | 4 | 4 | 4 |
| 2 | 16 | 4 | 8 | 4 | 4 | 4 | 4 | 4 |
| 3 | 16 | 4 | 8 | 4 | 8 | 4 | 4 | 1 |
| 4 | 16 | 4 | 8 | 4 | 8 | 2 | 4 | 1 |

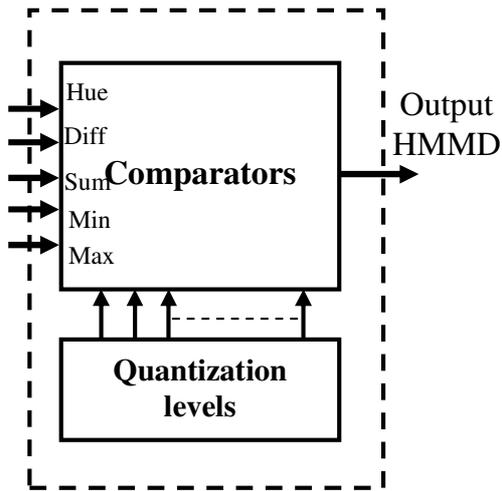

Figure 5    Block quantization

## 3.3. Presentation of our solution for the CSD hardware implementation

This solution is based on the association of several blocks RAM. Instead of using only one memory to stock the frame, in this solution, we used 10 BRAMS. We used 10 structures 8 * 8 sweeping the 10 BRAMS at the same time, so this method promotes the parallel processing. Through this method of execution time is divided by 10. Video frame data are multiplexed and sent to the BRAMS as is explained in figure 7 and figure 8

This work is motivated by the existence of a sufficient number of "Brams" on the FPGA Virtex 2 Pro. The organigram of the video addresses Ram's control is explained in Figure 9.

Wish Bram stores 10*8 pixels, so to scan one picture we use two horizontal shifts for every Bram. This method can accelerate the process and makes possible real time implementation of this descriptor.

This work is a prototype phase for the implementation of such applications in ASICs and integration into new generations of communication devices such as mobile phones

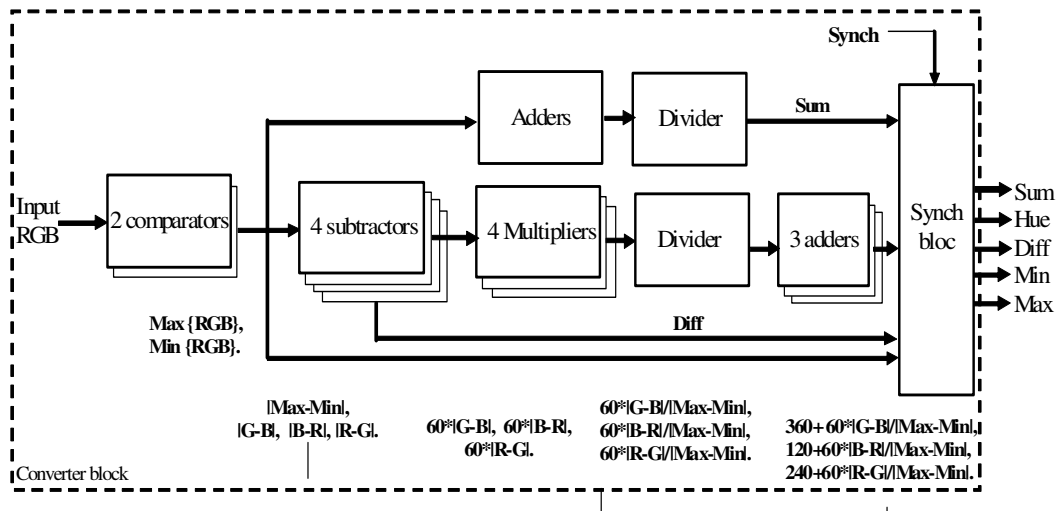

Figure 6    Bloc diagram of RGB to HMMD conversion

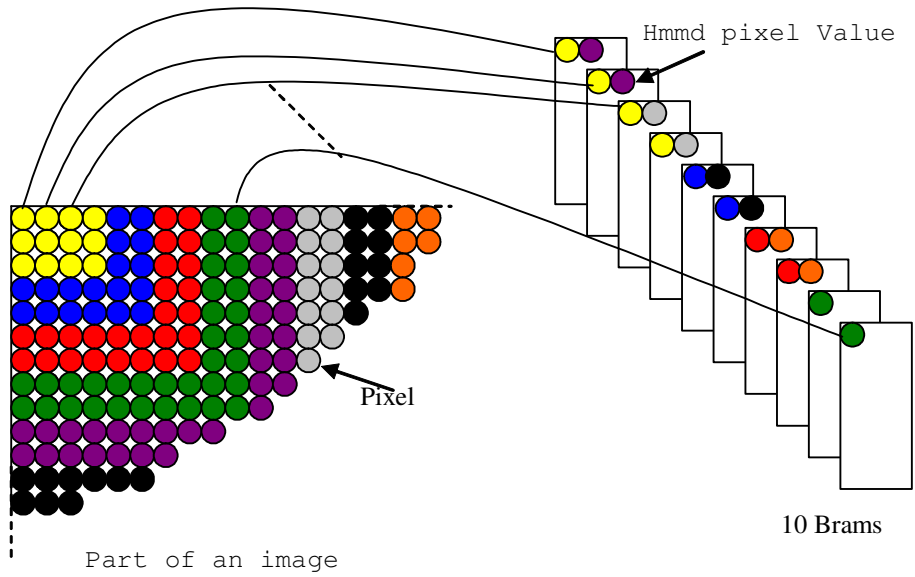

Figure 7 The principle of this method

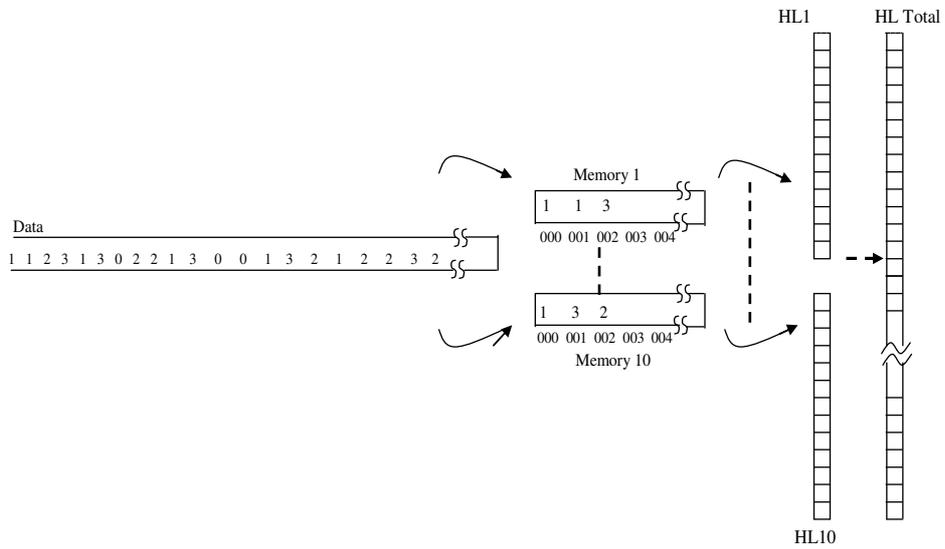

Figure 8 Video data distribution

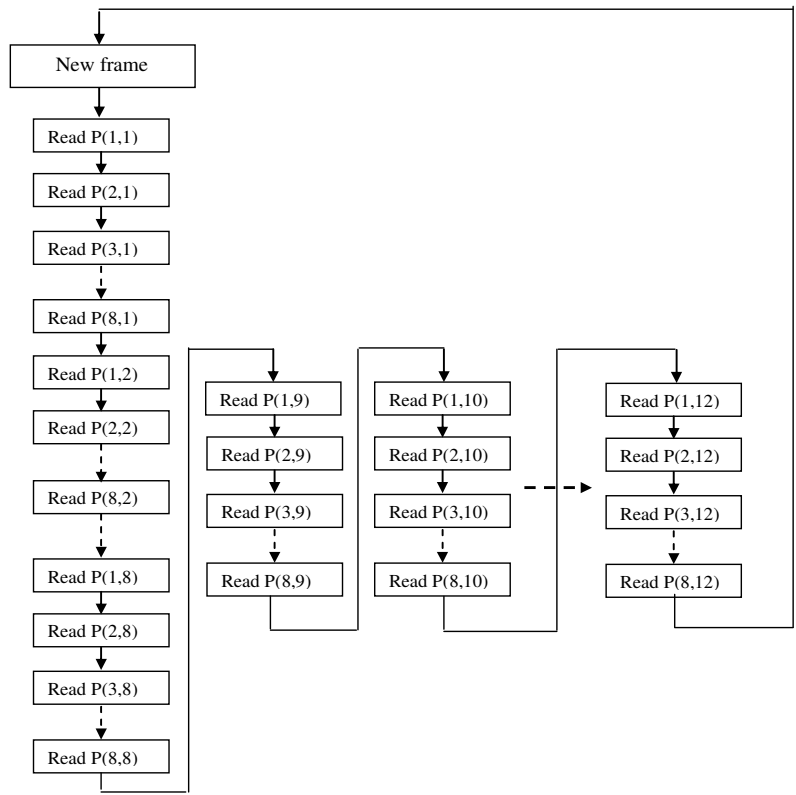

Figure 9 Addresses RAM Generation

## 4  EXPERIMENTAL RESULTS

In this section, we demonstrate the effectiveness of the hardware system for video processing application. The input video can either be obtained through a camera system or other video devices. Figure 10 (a) shows the min and max calculation from real video processing image and Figure 10 (b) shows a snapshot of the system logic and timing simulation of the output Hue calculation. The Figure 10 (c) shows a snapshot of the timing simulation of rhe RGB/HMMD block conversion.

Figure 11 shows a snapshot of the demultiplexor. The role of this block is to distribute data to partials memories BRAMS within the constraints of real time. The simulation is performed by the ModelSim tool

The RTL synthesis of the Brams association is given by the figure 12. The following figure shows a Simplified schematic of the distribution of the embedded memory blocks

Figure 13 shows the block diagram of the proposed method

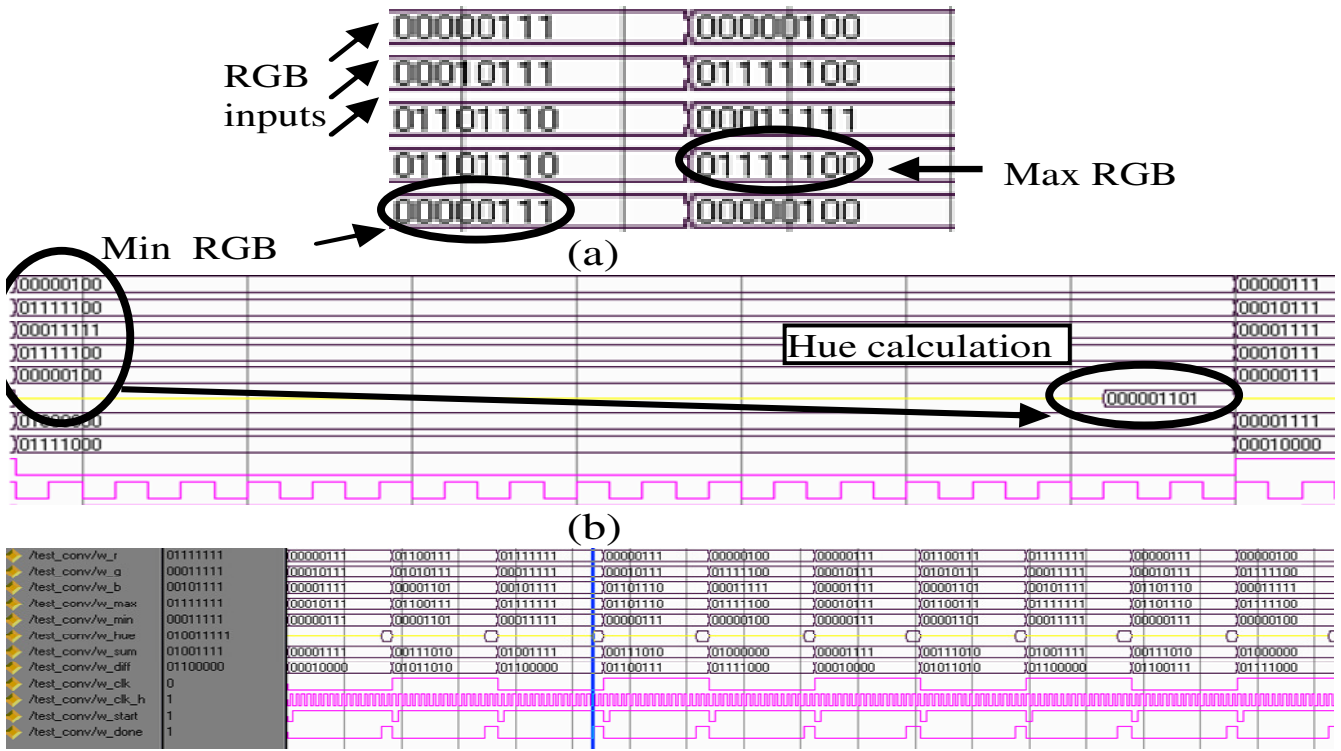

(a)

(b)

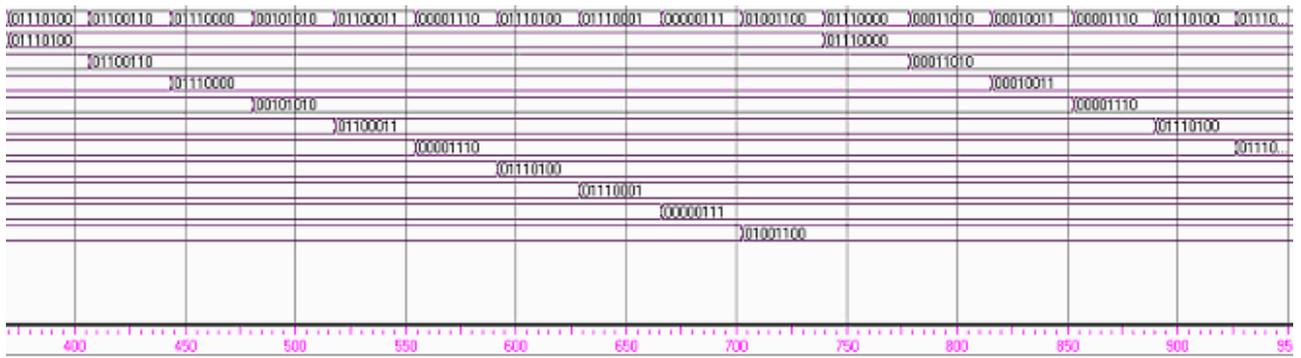

Figure 10 (c) Process of the HMMD

Figure 11 Process of the demultiplexor

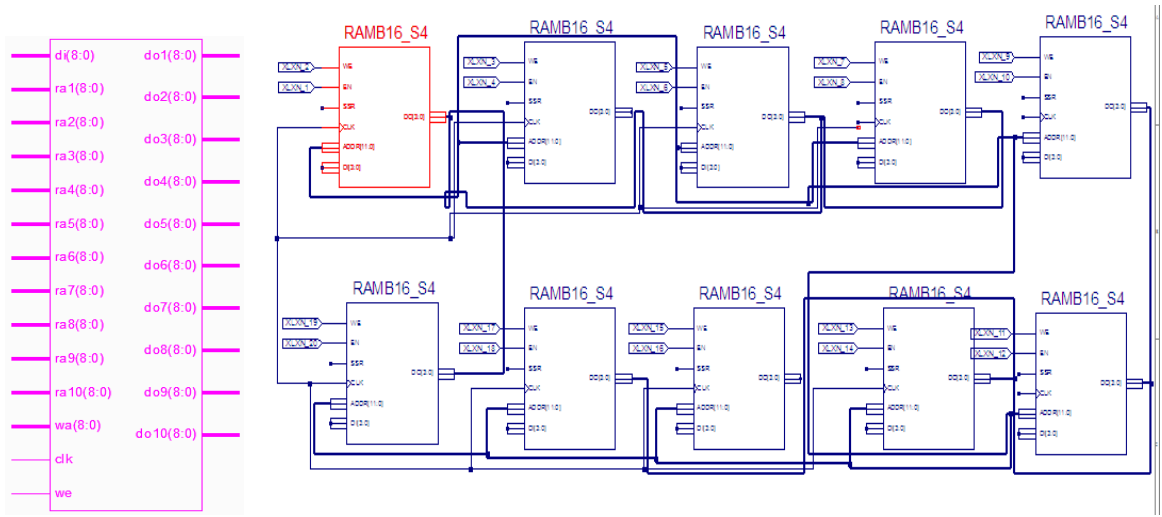

Figure 12 Block diagram of the Brams association

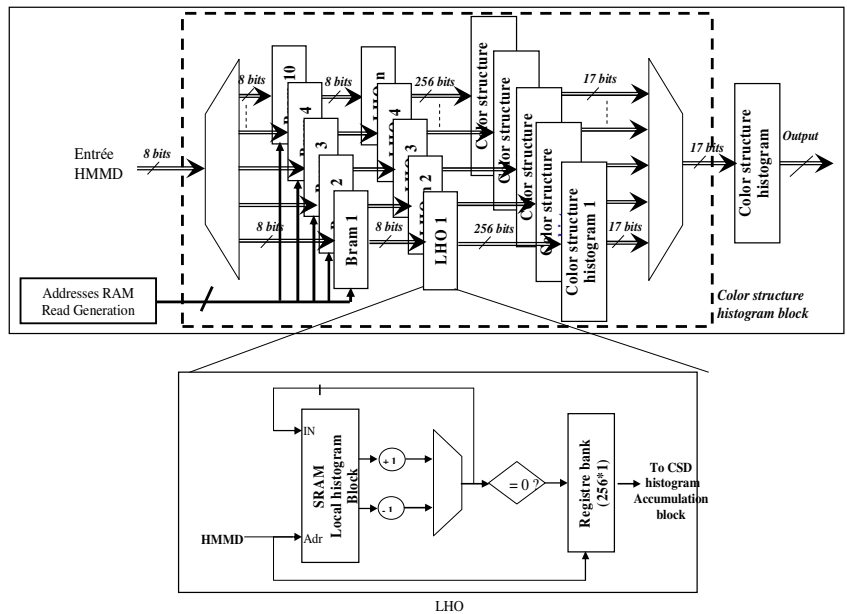

Figure 13 The principle of the proposed method

The design is synthesized for Xilinx-II Pro (XC2VP30) – see Figure 14 This device contains 30,816 logic cells, two PowerPc 4O5 (PPC) embedded processors [12,13]. The FPGA is situated on Xilinx XUP Virtex-II Pro development board, which also contains a configuration PROM and various useful interfaces [14, 15], 2448kb of block RAM,. The DDR SDRAM DIMM can support up to 2Gbytes of RAM. This board has useful interface ports, RS-232 port, and others. It also has various expansion connectors to expand the usability of this board to meet the requirements of different video and image processing applications. Our major purpose of this system is to implement the entire hardware platform to provide a general solution for video and image processing, and demonstrate its effectiveness through various application scenarios.

TABLE 3.
RESULTS OF THE SYNTHESIS

| Resources | Used |
|---|---|
| Slices | 183 out of 13696 |
| Adders/Subtractors | 5 |
| Latches | 16 |
| Comparators | 397 |
| Brams | 76.8 of 2448Kbits |
| Multipliers | 2 out of 136 |
| LUTs | 354 out of 27,392 |
| GCLK | 1 out of 16 |

The Integrated Software Environment (ISE) gave the results presented in Table 3. We notice that the occupied area of the design is almost 1% of total resources. The minimum period is 8.887 ns, and has total power consumption around 115mw.

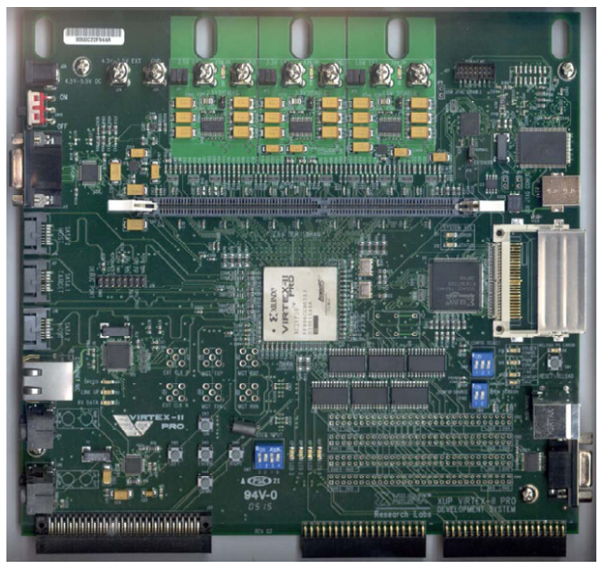

Figure 14 The prtotype board with the Virtex-II Pro.
Source: Xilinx

To evaluate the color structure descriptor we have used a rich corpus of different image genres. An example of retrieval results using the CSD with 256 quantification levels is given in figure 15. The obtained results demonstrate the performance of the CSD for color based image retrieval in 256 or quantification levels.

## 5 CONCLUSION

FPGAs are often used as implementation platforms for real-time image processing applications because their structure can exploit spatial and temporal parallelism. Using high-level languages and compilers to hide the constraints and automatically extract parallelism from the code does not always produce an efficient mapping to hardware. The code is usually adapted from a software implementation and thus has the disadvantage that the resulting implementation is based fundamentally on a serial algorithm.

Global operations such as chain coding require random access to memory and cannot be easily implemented under stream processing modes. This forces the designer to reformulate the algorithm. In this paper, an architecture which can generate CSD description image at 25 fps is proposed. Detailed design explorations of the hardware implementation. We provide the vision of future MPEG-7 descriptor applications for not only indexing and retrieval, but also for real-time multimedia applications.

Use 10 BRAMS instead of a single block RAM, accelerated the process because it promotes the parallel treatment. And we will prove the superiority of the proposed system. Through this, we expect that it will be widely used for applications such as digital TV and so on.

In the future work, it would be interesting to integrate more complicated video processing modules into this platform like Telecom and Video applications in real time

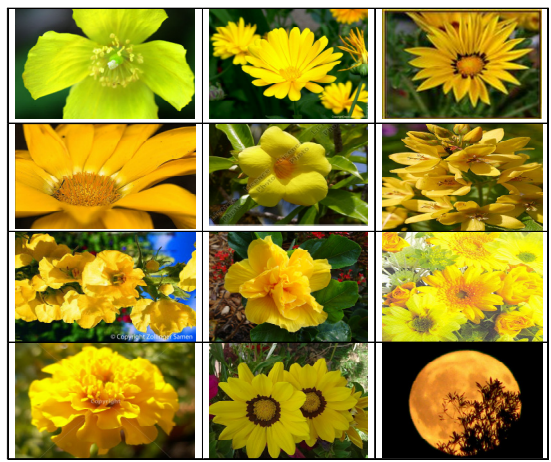

Figure 13. Example of image retrieval results with the CSD descriptor – 256 quantification levels

## REFERENCES

[1] C.T. Johnston, K.T. Gribbon, D.G. Bailey, "Implementing Image Processing Algorithms on FPGAs", Proceedings of the Eleventh Electronics New Zealand Conference, ENZCon'04, Palmerston North,

[2] T Morimoto, H Adachi, K Yamaoka "Image-Scan Video Segmentation Architecture Based on embedded Memory Technology",. Research Center for Nanodevices and systems, Hiroshima University Japan.

[3] C. Christopoulos, D. Berg and Athanassios "The colour in the upcoming MPEG-7 standard"

[4] Michael Papazachariou: An Overview of MPEG-7 Colour Descriptors. ICBR Journal Club, 29 Mars, 2004, http://www.cs.bris.ac.uk/home/janko/journalclub/michael1.pdf.

[5] Messing, D.S, van Beek. P, Errico. J.H., "The MPEG-7 colour structure descriptor: image description using colour and local spatial information", International Conference on Image Processing, Thessaloniki, Greece, 2001, ISBN: 0-7803-6725-1.

[6] Jens-Rainer Ohm, Heon Jun Kim, Santhana Krishnamachari, B. S. Manjunath, Dean S. Messing, Akio Yamada "The MPEG-7 Color Descriptors".

[7] Jing-Ying Chang, Chung-Jr Lian, Hung-Chi Fang, Liang-Gee Chen, "Architecture and Analysis of Color Structure Descriptor for Real-Time Video Indexing and Retrieval", Advances in Multimedia Information Processing - PCM 2004, LNCS 3332, pp. 130-137, 2004.

[8] Messing, D.S, van Beek. P, Errico. J.H., "The MPEG-7 colour

structure descriptor: image description using colour and local spatial information", International Conference on Image Processing, Thessaloniki, Greece, 2001, ISBN: 0-7803-6725-1.

[9]  B. S. Manjunath, Jens-Rainer Ohm, Vinod V. Vasudevan and Akio Yamada: Color and Texture Descriptors. IEEE Transactions on Circuits and Systems for Video Technoligiy, Vol. 11, NO. 6, pp. 703-715, June 2001

[10]  W.Stechele: Video Processing using Reconfigurable Hardware Acceleration for Driver Assistance. Workshop on Future Trends in Automotive Electronics and Tool Integration at DATE 2006, Munich, March 6-10, 2006.

[11]  ISO/IEC JTC1/SC29/WGI1, Doc N3913: Study of CD 15938-3 MPEG-7 Multimedia Content Description Interface – Part 3 Visual. ISO/IEC January 2001 (Pisa).

[12]  Virtex-II Pro and Virtex-II Pro X Platform FPGAs: Complete Data Sheet, Product Specification DS083 (v4.0) June 30, 2004.

[13]  XUP Virtex-II Pro Development System, November 9, 2004.

[14]  Hardware Reference Manual: Xilinx University Program Virtex-II Pro Development System UG069 (v1.0), March 8, 2005.

[15]  Diligent Video Decoder Board (VDEC1) Reference Manual, Revision: December 4, 2005

## ACKNOWLEDGEMENTS

I express my appreciation to Abdellatif Mtibaa for their comments and suggestions that improved the style of this paper, and also for their extensive hard work and sincere support in completing this project.

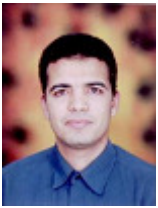

**Lamjed Touil**      Received his degree in Electrical Engineering from ENIM (National Engineering School of Monastir) in 2002 and his DEA in Electronics and Microelectronics degree from FSM (Science Faculty of Monastir) in 2006. In 2003 he joined the high Institute of technology of Sousse (ISET), Tunisia, as Technologue. he is a member of the Monastir laboratory of Electronics and microelectronics. His current research interests include FPGA and hardware implementation of video processing applications

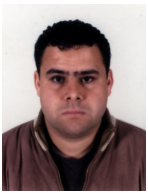

**Abdessalem Ben Abdelali**   Received his degree in Electrical Engineering and his DEA in Electronic Engineering from Sfax National School of Engineering (ENIS), Tunisia, in 2002. He received his Ph.D from (ENIS) and Bourgogne University (UB), France, in 2007. Since 2007 he has been joined high Institute of Informatics and Mathématic (ISIM), Monastir, Tunisia.

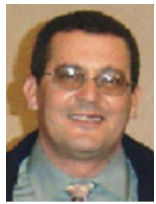

**Abdellatif Mtibaa** Received his PhD degree in Electrical Engineering at the National School of Engineering of Tunis. Since 1990 he has been an Assistant Professor in Micro-Electronics and Hardware Design with Electrical Department at the National School of Engineering of Monastir. Since 2007, he has been a professor at the electrical engineering department at the ENIM. His research interests include high level synthesis, rapid prototyping and reconfigurable architecture for real-time multimedia applications.

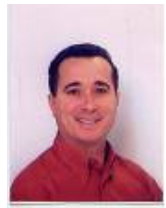

**El-Bay Bourennane**       Received his degree in Telecommunications Engineering from Sétif University (Algérie) in 1990, and he received his DEA in Signal Processing Image and Parole from Grenoble school of Engineering and Phd training (INPG). He received his Ph.D. in image processing from Bourgogne University (UB), France. He is currently a Professor in Bourgogne University.